# $PM_{2.5}$-GNN: A Domain Knowledge Enhanced Graph Neural Network For $PM_{2.5}$ Forecasting

Shuo Wang[*]
shawnwang.tech@gmail.com
School of Systems Science, Beijing
Normal University
Beijing, China

Yanran Li[*]
csyli@comp.polyu.edu.hk
Department of Computing, The Hong
Kong Polytechnic University
Hong Kong, China

Jiang Zhang[†]
zhangjiang@bnu.edu.cn
School of Systems Science, Beijing
Normal University
Beijing, China

Qingye Meng
mengqingye@caiyunapp.com
ColorfulClouds Pacific Technology
Co., Ltd.
Beijing, China

Lingwei Meng
lingweim@princeton.edu
Atmospheric and Oceanic Sciences
Program, Princeton University
Princeton, NJ, USA

Fei Gao
Philip.sss@mail.bnu.edu.cn
School of Systems Science, Beijing
Normal University
Beijing, China

## ABSTRACT

When predicting $PM_{2.5}$ concentrations, it is necessary to consider complex information sources since the concentrations are influenced by various factors within a long period. In this paper, we identify a set of critical domain knowledge for $PM_{2.5}$ forecasting and develop a novel graph based model, $PM_{2.5}$-GNN, being capable of capturing long-term dependencies. On a real-world dataset, we validate the effectiveness of the proposed model and examine its abilities of capturing both fine-grained and long-term influences in $PM_{2.5}$ process. The proposed $PM_{2.5}$-GNN has also been deployed online to provide free forecasting service.

[*]Both authors contributed equally to this paper.
[†]Jiang Zhang is the correspondence author of this paper.

## 1 INTRODUCTION

In the past few decades, the rapid development of industry has caused severe air pollution problem, especially in city clusters like the Beijing-Tianjin-Hebei, Yangtze River Delta and Sichuan Basin. In particular, particles smaller than 2.5μm ($PM_{2.5}$) have been paid special attention as they play an important role in atmospheric visibility reduction, human health, acid deposition and the climate [17].

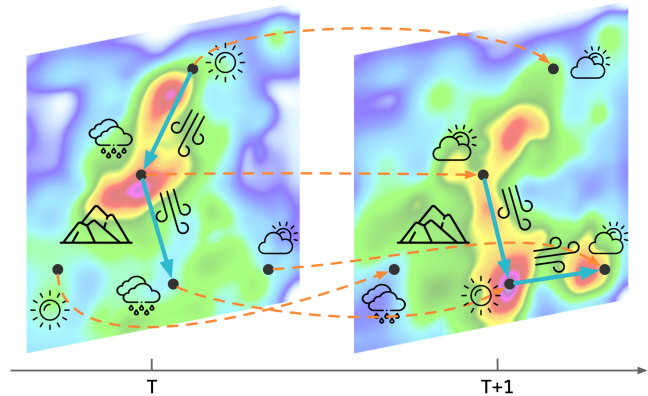

**Figure 1: A demonstration of $PM_{2.5}$ characteristics. Colors on the maps indicate the $PM_{2.5}$ concentrations. Directed blue lines represent transport from one area to the other, while dashed lines linking the same area denote diffusion along timesteps. Meteorological information (weather) and geographical knowledge (mountains) are also shown.**

However, it is non-trivial to accurately predict $PM_{2.5}$ concentration for several reasons. Firstly, $PM_{2.5}$ concentration is characterized by a complex process, starting from emission generated by pollution sources to transport and diffusion influenced by meteorological and geographical information [29]. It is thus a necessity to make good use of domain knowledge when modeling this temporal and spatial process. Secondly, these aforementioned $PM_{2.5}$ factors often take a wide-range and long-lasting effects. As reported by [6], $PM_{2.5}$ can be transported hundreds of kilometers in 72 hours. The ability of handling long-term dependencies is then another key to prediction accuracy. We summarize these $PM_{2.5}$ characteristics as: (1) **domain-knowledge sensitivity**; and (2) **long-term dependency**, which are demonstrated in Figure 1.

Existing work often structure $PM_{2.5}$ data into graphs and adopt graph-based approaches to capture the process. [10] establishes an undirected graph and computes the spatial similarity by utilizing the neighborhood information within a single city. [16] also operates on an undirected graph and derives the hidden spatial dependencies mainly based on a Graph Convolution Network. Nevertheless, none of these approaches is capable of explicitly incorporating domain knowledge like wind directions, which are critical for modeling $PM_{2.5}$ transport process. Partially due to this kind of incapability, these existing approaches fail to predict $PM_{2.5}$ concentration in a wide range for a long period of time.

In this work, we develop $PM_{2.5}$-GNN, which takes into account the aforementioned two $PM_{2.5}$ characteristics. Notably, we establish a *directed* graph where nodes are cities and edges representing city-to-city interactions. Since $PM_{2.5}$ prediction is sensitive to domain knowledge, we innovatively consider both meteorological and geographical information to precisely model its temporal and spatial process. The meteorological information for each city is characterized as node and edge features, while geographical knowledge among cities are encoded into graph structures. Based on the constructed graph, we learn $PM_{2.5}$ spatial transport among cities by leveraging a knowledge-enhanced Graph Neural Network, and capture the temporal diffusion process using a Recurrent Neural Network. Combing these two modules, the proposed $PM_{2.5}$-GNN is effective in modeling both transport and diffusion process by exploiting extensive domain knowledge. As a result, on a real-world dataset, we are able to accurately predict $PM_{2.5}$ concentrations of multiple city clusters for 72 hours in the future. This demonstrates the advantage of the proposed approach in characterizing $PM_{2.5}$'s long-term dependency.

In brief, we highlight our contributions as follows:

- We regard the necessity of domain knowledge in $PM_{2.5}$ prediction, and incorporate them into graph-structured data.
- We develop $PM_{2.5}$-GNN, a novel $PM_{2.5}$ prediction model to explicitly model the long-term dependency by utilizing domain knowledge.
- We introduce a large-scale real-world dataset, namely KnowAir, where we validate the effectiveness of our model through performing 72-hour $PM_{2.5}$ prediction. We release the dataset and code in hope to benefit future work in the research community.[1].
- We deploy $PM_{2.5}$-GNN in an online website as in Figure 2, and provide real-time $PM_{2.5}$ forecasting through free API access for third-party clients. Please refer to Appendix for more deployment details.

## 2 RELATED WORK
### 2.1 $PM_{2.5}$ Prediction
Recent literature on predicting $PM_{2.5}$ concentration often rely on deep learning based models [24, 28]. According to the input data structure, these work can be roughly classified into image-based and graph-based methods. For image-based methods, the basic idea is to interpolate the concentration numbers from monitoring stations to an image, and then use Convolutional Neural Networks (CNNs) to model the spatial correlations between the grids of the image [1, 13]. However, the monitoring stations are not uniformly distributed in urban and rural areas, and CNNs are incompetent of distinguishing the urban and rural grids on the image. This inevitably leads to erroneous modeling of PM2.5 process and results in imprecise concentration prediction. To alleviate this issue, [28] separates the image into several regions and proposes a Spatial Transformation Component (STC) to convert grids from same regions into a consistent input to simulate the pollutant sources.

[1]https://github.com/shawnwang-tech/PM2.5-GNN

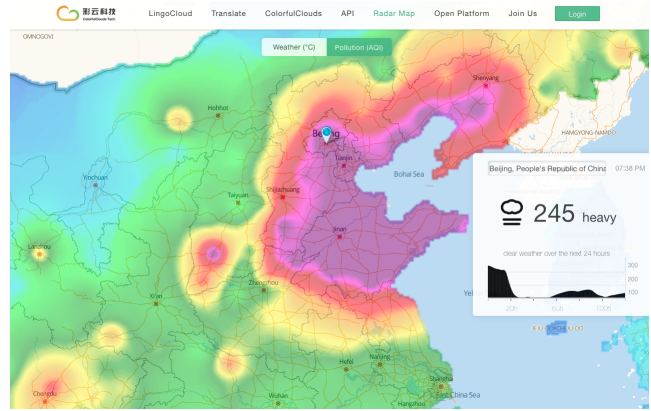

**Figure 2: Online website (http://caiyunapp.com/map/) that provides 72-hour real-time $PM_{2.5}$ concentration prediction using $PM_{2.5}$-GNN model proposed in this paper.**

In contrast, graph-based models naturally sidestep this issue since they shape the concentration numbers into graph nodes and keep their original distributions in graph structures. Along this line, GC-DCRNN [10] constructs the graph based on the environment similarity of the air sensors and captures the spatial dependency using a graph-structured diffusion convolution module. GC-LSTM [16] constructs graph where each node represents a monitoring station and devises a Graph Convolutional Network (GCN) to learn the spatial and temporal correlations among the graph series. Nevertheless, these approaches lack mechanisms of utilizing domain knowledge and thus only apply to $PM_{2.5}$ forecasting in a relatively small region within a short period. Our work is distinct from previous work in that by leveraging Graph Neural Networks, we are able to utilize extensive domain knowledge and model the complex $PM_{2.5}$ process more accurately.

### 2.2 Graph Neural Network
There has been an increasing attention on Graph Neural Networks (GNNs) due to their ability of extracting graph topological information for downstream tasks [30, 31]. Basically, GNNs allow nodes information to communicate along the edge paths on the graph. [27] generalizes this idea and establishes the message passing scheme for GNN computation where nodes representation are recursively updated by aggregating their neighborhood information. In one iteration, each node aggregates all the feature vectors of their $k$-ordered ($k$ is pre-defined) neighbors according to the graph structure. Typically, aggregation methods are differentiable functions, such as sum, mean, max or attention-based methods [21]. Without loss of generalization, by operating this kind of message passing scheme on directed graphs, features of source nodes will be passed to the target nodes through their directed edges.

Considering their flexibility and generalization, GNNs are especially effective for graph-structured problems when combining with other deep learning modules. In this work, we innovatatively apply GNNs to PM$_{2.5}$ concentration prediction. By integrating with a Recurrent Neural Network, our PM$_{2.5}$-GNN is able to model both transport and diffusion process in a long term within a wide range. In comparison with previous approaches merely being tested on small-scale datasets [10, 16], our work examines the ability of air forecasting on a large-scale real-world dataset, KnowAir, which will benefit both research and industry communities.

## 3 METHODOLOGY

In this section, we firstly define the PM$_{2.5}$ concentration prediction problem in mathematical formula. Then, we describe the data we used in this work and construct the graph consisting of the studied cities and their potential interactions. Based on the graph, we finally present the proposed approach to model the spatial and temporal dependencies of PM$_{2.5}$.

### 3.1 Studied Area and Data

To examine the models' abilities of capturing long-term dependencies, we select a wide range of areas ($103°E - 122°E$ and $28°N - 42°N$) as our study subject, that covers several severely polluted regions in China. Note this studied area is far large from those in previous work [16]. Please refer to Appendix to find the studied area shown on the map.

Following previous work [28], we use two types of data: PM$_{2.5}$ concentrations and weather forecasting data. PM$_{2.5}$ concentrations are obtained from Ministry of Ecology and Environment (MEE).[2] For weather forecasting data, we adopt two sources to fit online and offline systems differently. For online deployment, we acquire weather data from Global Forecast System (GFS)[3] since it is charge-free. GFS publishes its forecasting 4 times a day with few time delays, which contains future 15-days weather data with detailed information (which will be introduced in Section 3.3). For offline system, we obtain climate reanalysis ERA5[4] from European Centre for Medium-Range Weather Forecasts (ECMWF)[5] as the weather forecasting data. In addition to these two types of data, we also exploit domain knowledge to empower the model, which will be described in Section 3.3.

### 3.2 Problem Definition

Typically, PM$_{2.5}$ concentration prediction is formulated as a spatio-temporal sequence prediction problem. Let $X^t \in \mathbb{R}^{N \times 1}$ denote the PM$_{2.5}$ concentrations at time step $t$, where $N$ is the number of nodes. We define a *directed* graph $G = (V, E)$, where $V$ is the set of nodes representing cities, and $E$ is the set of edges denoting potential interactions among cities. Let $P^t \in \mathbb{R}^{N \times p}$ and $Q^t \in \mathbb{R}^{M \times q}$ denote the nodes' and edges' attribute matrices respectively at time step $t$, where $p$, $q$ are the corresponding attribute numbers, and $M = |E|$ is the edge size.

To enhance model's prediction ability, the key centers in how to better utilize domain knowledge. Our idea is to explicitly encode domain knowledge, e.g., meteorological information, into the attribute matrices and graph structures. Following previous work [28], we also exploit domain knowledge from the near future, which can be acquired from weather forecasting services. Formally, for any starting point $t$, we feed the observed PM$_{2.5}$ concentrations $X^t$ at current time $t$, next $T$ steps of attribute matrices $[P^{t+1}, \ldots, P^{t+T}]$ and $[Q^{t+1}, \ldots, Q^{t+T}]$, as well as the graph structure $G$ into our model. In this way, the PM$_{2.5}$ concentration problem is framed as:

$$\left[X^t; P^{t+1}, \ldots, P^{t+T}; Q^{t+1}, \ldots, Q^{t+T}; G\right] \xrightarrow{f(\cdot)} \left[\hat{X}^{t+1}, \ldots, \hat{X}^{t+T}\right] \quad (1)$$

As illustrated in Figure 3, we achieve this by iterating $T$ steps of the proposed model $PM_{2.5}$-GNN, denoted as $g(\cdot)$ in Equation 3. And $X^t$ is used at $t = 0$ when previous predictions are unknown. $\Theta$ stands for model parameters.

$$f(\cdot) = \underbrace{g(\ldots g(g(\cdot)))}_{T \text{ times}} \quad (2)$$

where

$$\hat{X}^{t+\tau} = \begin{cases} g(\hat{X}^{t+\tau-1}; P^{t+\tau}, Q^{t+\tau}, G, \Theta), & \forall \tau \in [1, \ldots, T] \\ X^t, & \tau = 0 \end{cases} \quad (3)$$

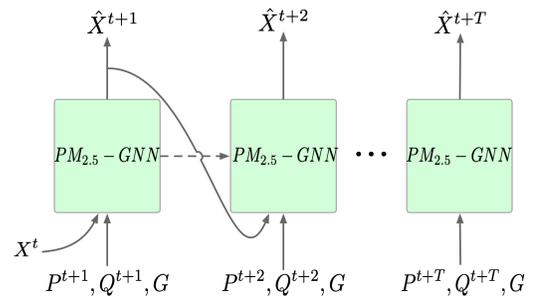

Figure 3: Illustration of PM$_{2.5}$-GNN model's process.

For training, we are to minimize the error between predicted values $[\hat{X}^1, \ldots, \hat{X}^T]$ and ground truth $[X^1, \ldots, X^T]$ using Mean Squared Error (MSE) loss:

$$\text{MSE Loss} = \frac{1}{T} \sum_{t=1}^{T} \left( \frac{1}{N} \sum_{i=1}^{N} \left( \hat{X}_i^t - X_i^t \right)^2 \right) \quad (4)$$

---

[2] http://datacenter.mee.gov.cn/websjzx/dataproduct/airproduct/queryAirAround.vm
[3] https://www.ncdc.noaa.gov/data-access/model-data/model-datasets/global-forcast-system-gfs
[4] https://climate.copernicus.eu/climate-reanalysis
[5] https://www.ecmwf.int/

Table 1: Meteorological Attributes of Nodes ($P$)

| Variable Name | Unit |
| --- | --- |
| Planetary Boundary Layer (PBL) height | $m$ |
| K index | $K$ |
| u-component of wind | $m/s$ |
| v-component of wind | $m/s$ |
| 2m Temperature | $K$ |
| Relative humidity | $\%$ |
| Total precipitation | $m$ |
| Surface pressure | $Pa$ |

Table 2: Attributes of Edges ($Q$)

| Variable Name | Unit |
| --- | --- |
| Wind speed of source node $|v|$ | $km/h$ |
| Distance between source and sink $d$ | $km$ |
| Wind direction of source node $\beta$ | (°) |
| Direction from source to sink $\gamma$ | (°) |
| Advection coefficient $S$ (Equation 5) | $\%$ |

## 3.3 Graph Construction

As analyzed before, $PM_{2.5}$ is influenced by a number of factors. To improve prediction accuracy, we establish our graph by incorporating domain knowledge as attributes on nodes and edges, then learn transport and diffusion process based on the graph.

**Node Attributes**. A place's meteorological characteristics influence how the pollutant will be diffused. Following are the representative meteorological variables we choose as nodes' attributes:

- **Planetary Boundary Layer (PBL) height**: PBL height is related to pollutants' vertical dilution [18]. As shown by [12], PM2.5 concentration has an inversely linear correlation with PBL height.
- **K index**: The stratification instability of moist air also influences $PM_{2.5}$ concentration, which can be represented by K index. A higher K index indicates that the tropospheric stratification is more unstable, and the fog and haze is often weaker [17].
- **Wind speed**: Studies [6] show that there exists an inverse correlation between $PM_{2.5}$ and wind speed. Since $PM_{2.5}$ concentration often peaks at 500 meters high, we introduce two node attributes representing the wind speed at this level (950 hPa) in two directions.
- **2m Temperature**: Temperature is another factor influencing $PM_{2.5}$ concentration through chemical effects and ventilation of cold fronts [19].
- **Relative humidity**: As indicated by [19], high surface relative humidity promotes the formation of ammonium nitrate, which is a component of $PM_{2.5}$.
- **Precipitation**: It is revealed that the drag effect of precipitation produces wet scavenging and downdrafts in near-surface atmosphere, resulting in a negative correlation between precipitation and $PM_{2.5}$ concentration [15, 17].
- **Surface pressure**: As shown by [9], pressure presents a significantly positive coherence with $PM_{2.5}$.

These node attributes are related to $PM_{2.5}$ vertical diffusion through dynamic or thermodynamic effects. We summarize them in Table 1. Besides, we also use hour and day of the week (exact time) as temporal information like [16].

**Edge Attributes**. [6] shows that wind speed and direction have a decisive effect on $PM_{2.5}$ horizontal transport. To incorporate this strong domain knowledge, we develop a simplified model to capture transport from one place to another. As illustrated in Figure 4, the start place is the $PM_{2.5}$ source represented as a source node $j$, and the target place is denoted as a sink node $i$. We include several variables related to these two nodes into the model and get inspired from [32] to estimate the amount of $PM_{2.5}$ transported from $j$ to $i$.

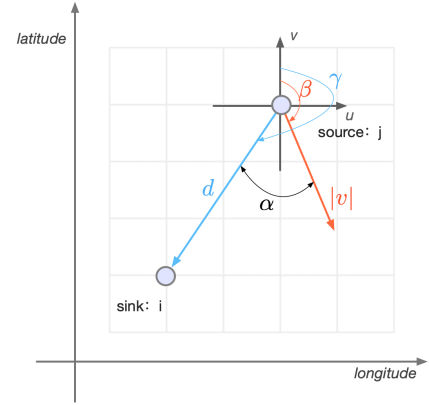

Figure 4: A simplified model for transport estimation.

As shown in Equation 5, we derive an advection coefficient $S$ as the wind impact degree on sink node by the source node, where $|v|$ is the wind speed of the source $j$, $d$ is the distance between $j$ and $i$, and $\alpha$ is the angle between two directions: one is the wind direction $\beta$ of the source, and the other one is the edge direction $\gamma$ pointing from source to sink. In other words, $\alpha = |\gamma - \beta|$. One thing to note is that we adopt ReLU function to merely capture the uni-directional transport from source to target place. Here, the direction of edge on source node is identical to the wind direction following the meteorological convention.

$$S = \text{ReLU}\left(\frac{|v|}{d}\cos(\alpha)\right) \quad (5)$$

In total, we summarize the features in the edge attribute matrix $Q$ in Table 2.

**Adjacency Matrix**. In addition, we need to compute the nodes' correlations to build the adjacency matrix. Intuitively, most aerosol pollutants are distributed in a certain range above the ground [22]. Moreover, the mountains lying along the two cities will hinder $PM_{2.5}$ transport of the pollutants. Based on these intuitions, we constrain the weights in the adjacency matrix through:

$$A_{ij} = H(d_\theta - d_{ij}) \cdot H(m_\theta - m_{ij}), \quad \text{where}$$
$$d_{ij} = ||\rho_i - \rho_j||$$
$$m_{ij} = \sup_{\lambda \in (0,1)} \{h(\lambda \rho_i + (1-\lambda)\rho_j) - \max\{h(\rho_i), h(\rho_j)\}\} \quad (6)$$

where $\rho_i$ is the position (latitude, longitude) of node $i$, $h(\rho)$ is the altitude of the position $\rho$, and $||\cdot||$ is the L2-norm of a vector. $H(\cdot)$ is the Heaviside step function,[6] where $H(x) = 1$ if and only if $x > 0$. $d_\theta$ and $m_\theta$ are the thresholds of distance and altitude, respectively. In this paper, we set $d_\theta = 300km$ for the distance threshold and $m_\theta = 1200m$ for the altitude threshold. That means, PM$_{2.5}$ can transport from one city to another only if the distance between these two cities is less than $300km$ and the mountains between them are lower than $1200m$.

After incorporating domain knowledge into node-, edge-attributes and adjacency matrix, we construct a knowledge-aware PM$_{2.5}$ graph for concentration prediction. Note there are three distinctions in our graph. Firstly, we take 3 hours as the time interval for the task. Secondly, we choose prefecture-level cities as the graph's nodes. Position and PM$_{2.5}$ concentration of a city are calculated by averaging the corresponding numbers of the monitoring stations within the city. This allows us to focus on PM$_{2.5}$'s long-term dependency between cities. Thirdly, unlike previous work [16], the edges in our graph are directed. By capturing these detailed information, we finally succeed in predicting the PM$_{2.5}$ concentrations for future 72 hours.

### 3.4 PM$_{2.5}$-GNN Model

As illustrated in Figure 5, the proposed PM$_{2.5}$-GNN model consists of two main components. A knowledge-enhanced Graph Neural Network (GNN) is devised to capture pollutants' horizontal transport by leveraging neighboring information and updating nodes' representations. A spatio-temporal GRU is applied after updates to model pollutants' vertical accumulation and diffusion under the influence of weather.

Following the paradigm of message passing [4, 27], the knowledge-enhanced GNN learns representations by iteratively aggregating neighboring information on the graph. This iterative process is formulated as in Equation 7, where $\Psi$ and $\Phi$ are differentiable functions. At each time step, a node's representation $\xi_i^t$ is obtained by concatenating the previous predicted PM$_{2.5}$ concentrations $\hat{X}_i^{t-1}$ and its current attributes $P_i^t$. An edge's representation $e_{j \to i}^t$ is calculated by combining its connecting nodes as well as its own attributes.

$$\begin{aligned}
\xi_i^t &= [\hat{X}_i^{t-1}, P_i^t] & \forall i \in V \\
e_{j \to i}^t &= \Psi([\xi_j^t, \xi_i^t, Q_{j \to i}^t]) & \forall \langle j, i \rangle \in E \\
\zeta_i^t &= \Phi(\sum_{j \in N(i)} (e_{j \to i}^t - e_{i \to j}^t)) & \forall i \in V
\end{aligned} \quad (7)$$

Note that under our formulation, edge attributes $Q_{j \to i}^t$ and edge representations $e_{j \to i}^t$ are direction-aware. This kind of formulation enables us to explicitly measure the influence between node neighbors. Given a node $i$, the transport from its neighbor $j$ approximates to the difference between the import $e_{j \to i}^t$ and export influence $e_{i \to j}^t$, depicted as the orange and blue lines in Figure 5. Then, the spatial correlation $\zeta_i^t$ for node $i$ is calculated by summarizing the influences from all its neighbors. After several steps of recursion, each node on the graph is aware of others, and the transport is captured using the encoded domain knowledge.

---

[6]https://en.wikipedia.org/wiki/Heaviside_step_function

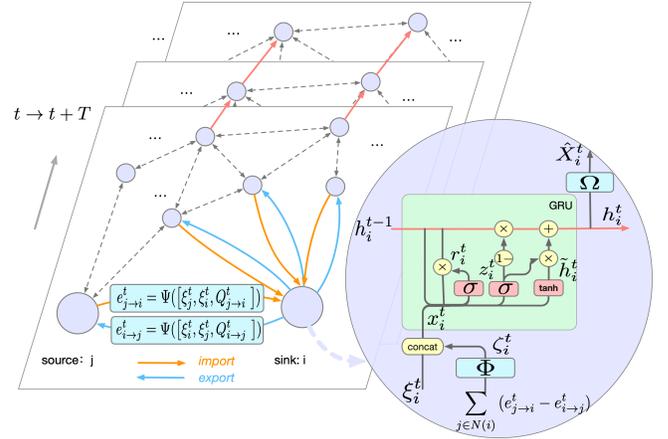

**Figure 5: The proposed PM$_{2.5}$-GNN model consisting of a knowledge-enhanced GNN and a spatio-tempo RNN.**

To learn the temporal diffusion of PM$_{2.5}$, we devise a Recurrent Neural Network (RNN) and apply it over the knowledge-enhanced GNN. In specific, we adopt Gated Recurrent Unit (GRU) [3] as our basic recurrent unit due to its advantage on capturing long-term dependency. At each time step, the GRU cell takes as input the combination of node's representation $\xi_i^t$ and its spatial correlation $\zeta_i^t$. This allows GRU to also consider the spatial transport while modeling the temporal diffusion. We depict the process of our spatio-tempo GRU in Equation 8:

$$\begin{aligned}
x_i^t &= [\xi_i^t, \zeta_i^t] \\
z_i^t &= \sigma\left(W_z \cdot [h_i^{t-1}, x_i^t]\right) \\
r_i^t &= \sigma\left(W_r \cdot [h_i^{t-1}, x_i^t]\right) \\
\tilde{h}_i^t &= \tanh\left(W \cdot [r_i^t * h_i^{t-1}, x_i^t]\right) \\
h_i^t &= (1 - z_i^t) * h_i^{t-1} + z_i^t * \tilde{h}_i^t
\end{aligned} \quad (8)$$

where $W_z$, $W_r$ and $W$ are the learnable parameters.

Based on the output of the proposed PM$_{2.5}$-GNN model, we finally predict PM$_{2.5}$ concentration by:

$$\hat{X}_i^t = \Omega(h_i^t) \quad \forall i \in V \quad (9)$$

where $\Omega$ is a Multi-Layer Perceptron (MLP). As summary, we outline the process of our PM$_{2.5}$-GNN model in Algorithm 1 in Appendix.

## 4 EXPERIMENTS

In this section, we conduct extensive experiments to demonstrate the effectiveness of PM$_{2.5}$-GNN. Then, we carefully analyze whether and to what extent PM$_{2.5}$-GNN is able to resemble PM$_{2.5}$'s characteristics of domain-knowledge sensitivity and long-term dependency. For reproducibility, we release the datasets and codes on GitHub. [7].

### 4.1 Datasets

Using the information described in Section 3.1, we construct a whole 4-year (2015/1/1 to 2018/12/31) dataset **KnowAir**, which covers in total 184 cities (nodes). To thoroughly investigate the model abilities, the dataset is further spilt into 3 ones.

As presented in Table 3, Dataset 1 is constructed by dividing the whole dataset into Train/Validate/Test sets with 2:1:1 in order to examine the models' air forecasting ability in a general long-range setting. Then, we follow [7] and develop Dataset 2, where we focus on heating season (November to February). Dataset 2 is more challenging for two reasons. Firstly, during winters, heating emissions can dramatically increase the frequency of air pollution occurrence [8]. Secondly, the direction of prevailing wind is north or northwest, which contributes to PM$_{2.5}$'s long-distance transport from North China to South China [25]. At last, we also provide Dataset 3, where information in past three months are used to predict PM$_{2.5}$ concentration in the current month. Note that Dataset 3 mocks the scenario how real-world online forecasting system utilizes the data and predict PM$_{2.5}$ concentrations.

Compared to previous KDD CUP of Fresh Air Dataset adopted by previous work [10], our dataset is large-scale in city coverage, and is accompanied by extensive knowledge. This fosters us to perform accurate real-time long-term (72 hours) PM$_{2.5}$ forecasting.

Table 3: Dataset is spilt into 3 sub-datasets.

| Dataset | Train | Validate | Test |
|---|---|---|---|
| 1 | 2015/1/1 - 2016/12/31 | 2017/1/1 - 2017/12/31 | 2018/1/1 - 2018/12/31 |
| 2 | 2015/11/1 - 2016/2/28 | 2016/11/1 - 2017/2/28 | 2017/11/1 - 2018/2/28 |
| 3 | 2016/9/1 - 2016/11/30 | 2016/12/1 - 2016/12/31 | 2017/1/1 - 2017/1/31 |

### 4.2 Compared Models

Our task is to predict the next 72 hours PM$_{2.5}$ concentrations given observed PM$_{2.5}$ concentrations at starting point and next 72 hours weather forecast data. To validate the effectiveness of the proposed PM$_{2.5}$-GNN, we compare with several models:

- **MLP** is a simple multi-layer neural network which takes as input the node representation at the current time step (same as Equation 7). It is the only compared model without a memory module.
- **LSTM** [5] consumes the sequence of node representations, and performs prediction over its last hidden state. In other words, LSTM solely models the temporal correlation between local weather condition and local PM$_{2.5}$.
- **GRU** [2] is similar to LSTM with a few difference in memory cells. By comparing with them, we aim at examining the importance of temporal information in PM$_{2.5}$ forecasting.

- **GC-LSTM** proposed by [16] is the current state-of-the-art baseline for PM$_{2.5}$ prediction. It integrates LSTM and Graph Convolutional Networks (GCN) to model the temporal and spatial dependency respectively. Different from our PM$_{2.5}$-GNN, the GCN module in GC-LSTM only applies to undirected graph, and no edges' attributes are used. These limit its capacity of explicitly modeling PM$_{2.5}$ transport along the direction of wind flow. For fair comparison, we also feed domain knowledge into it.
- **nodesFC-GRU** is a degrading version of the proposed PM$_{2.5}$-GNN. It is implemented by replacing the GNN module in PM$_{2.5}$-GNN with fully connected (FC) layers. We compare with it to examine how much improvement can be brought by neighborhood information using such a naive method. Since nodesFC-GRU has the largest perceptive field analogy to CNN [11], it can also be viewed as a rough comparison with CNN-based approaches. Its architecture details can be found in Appendix.

### 4.3 Experimental Settings

We run experiments on a single NVIDIA TITAN X. For pre-processing, we rescale the features of nodes and edges to have mean of 0 and standard deviation of 1 (unit variance). We initialize the GRU's hidden state $h^0$ with a zero tensor. Specifically, functions $\Psi$ and $\Phi$ in Equation 7 as well as $\Omega$ in Equation 9 are MLPs with layer numbers of 2, 1, 1, respectively. Moreover, the import and export influences in Equation 7 are implemented using pytorch_geometric (PyG) [4] and its dependent library pytorch_scatter.[8]

For fair comparison, we feed each compared model with identical input as well as domain knowledge. All models are trained using RMSprop [20] for 50 epochs with learning rate as $5^{-4}$. Early stopping is also adopted on the validation set. Following industrial practice, we rescale the model predictions into the range $0\mu g/m^3$ to $500\mu g/m^3$ using inverse standardization.

We evaluate the model performances using three groups of metrics: (1) train and test losses to show the model's ability of generalization; (2) mean absolute error (MAE) and root mean square error (RMSE) to directly examine the prediction accuracy; and (3) commonly used meteorological metrics [26] to measure the performance near the pollution threshold, including critical success index (CSI), probability of detection (POD), false alarm rate (FAR). Note that the higher CSI and POD are, the better the model performs.

In specific, MAE and RMSE are calculated using Equation 10:

$$RMSE = \sqrt{\frac{\sum_{i=1}^{n}(y_i - \hat{y}_i)^2}{n}}$$
$$MAE = \frac{\sum_{i=1}^{n}|y_i - \hat{y}_i|}{n} \quad (10)$$

where $y_i, \hat{y}_i$ are ground truth and the prediction, respectively, and $n$ is the total number of observed data samples. To calculate CSI, POD, FAR, we need to binarize the prediction and ground truth to 0/1 matrices using the threshold of PM$_{2.5}$ pollution indicating polluted or not. The threshold is chosen as $75\mu g/m^3$, the demarcation point of good air quality suggested by Ambient Air Quality Standards in China [14]. Then we calculate the hits (prediction = 1, truth = 1), misses (prediction = 0, truth = 1) and false alarms (prediction = 1, truth = 0). Finally, CSI, POD, FAR are reported using Equation 11:

$$\text{CSI} = \frac{\text{hits}}{\text{hits} + \text{misses} + \text{falsealarms}}$$
$$\text{FAR} = \frac{\text{falsealarms}}{\text{hits} + \text{falsealarms}} \quad (11)$$
$$\text{POD} = \frac{\text{hits}}{\text{hits} + \text{misses}}$$

---

[7] https://github.com/shawnwang-tech/PM2.5-GNN

[8] https://github.com/rusty1s/pytorch_scatter

Table 4: Overall performance on all three sub-datasets. Best scores are in bold.

| Dataset | Metric | MLP | LSTM | GRU | GC-LSTM | nodesFC-GRU | PM$_{2.5}$-GNN |
|---|---|---|---|---|---|---|---|
| 1 | Train_Loss | 0.5984 ± 0.0981 | 0.4424 ± 0.0399 | 0.4635 ± 0.0383 | 0.4157 ± 0.0484 | **0.3263 ± 0.0624** | 0.4692 ± 0.2853 |
|  | Validate_Loss | 0.5279 ± 0.0495 | 0.4514 ± 0.0136 | 0.4545 ± 0.0129 | 0.4455 ± 0.0125 | **0.4138 ± 0.0097** | 0.4158 ± 0.0177 |
|  | Test_Loss | 0.5647 ± 0.0454 | 0.4805 ± 0.0137 | 0.4830 ± 0.0124 | 0.4732 ± 0.0132 | 0.4395 ± 0.0088 | **0.4359 ± 0.0187** |
|  | RMSE | 22.98 ± 0.98 | 21.07 ± 0.38 | 21.13 ± 0.37 | 20.90 ± 0.40 | 20.28 ± 0.29 | **20.16 ± 0.48** |
|  | MAE | 18.37 ± 0.94 | 16.68 ± 0.39 | 16.77 ± 0.37 | 16.53 ± 0.41 | 15.98 ± 0.30 | **15.91 ± 0.49** |
|  | CSI | 40.77 ± 2.69% | 44.87 ± 1.09% | 44.71 ± 0.99% | 45.64 ± 1.10% | 47.61 ± 0.92% | **47.91 ± 1.65%** |
|  | POD | 51.43 ± 5.68% | 56.43 ± 2.43% | 56.17 ± 2.45% | 57.98 ± 2.51% | 59.79 ± 2.11% | **60.33 ± 3.42%** |
|  | FAR | 32.80 ± 4.29% | 31.21 ± 1.68% | 31.16 ± 1.80% | 31.65 ± 1.73% | **29.87 ± 1.43%** | 29.83 ± 2.36% |
| 2 | Train_Loss | 0.6094 ± 0.1291 | 0.4638 ± 0.0861 | 0.4808 ± 0.0806 | 0.4417 ± 0.0951 | **0.3368 ± 0.1227** | 0.4437 ± 0.0988 |
|  | Validate_Loss | 0.6356 ± 0.1113 | 0.5529 ± 0.0396 | 0.5465 ± 0.0359 | 0.5473 ± 0.0379 | 0.5575 ± 0.0339 | **0.5173 ± 0.0469** |
|  | Test_Loss | 0.6570 ± 0.0963 | 0.5825 ± 0.0350 | 0.5700 ± 0.0353 | 0.5750 ± 0.0413 | 0.5747 ± 0.0337 | **0.5375 ± 0.0494** |
|  | RMSE | 35.55 ± 2.76 | 33.53 ± 1.04 | 33.09 ± 1.00 | 33.20 ± 1.23 | 33.07 ± 1.03 | **32.11 ± 1.47** |
|  | MAE | 28.67 ± 2.52 | 26.90 ± 1.04 | 26.54 ± 0.97 | 26.57 ± 1.22 | 26.40 ± 0.97 | **25.68 ± 1.42** |
|  | CSI | 45.52 ± 5.49% | 49.75 ± 2.09% | 49.83 ± 1.79% | 50.13 ± 2.50% | 48.79 ± 1.38% | **51.35 ± 2.53%** |
|  | POD | 60.85 ± 9.17% | 64.94 ± 3.30% | 64.58 ± 3.03% | 64.54 ± 3.49% | 61.29 ± 2.07% | **66.24 ± 4.56%** |
|  | FAR | 34.56 ± 6.21% | 31.88 ± 2.28% | 31.31 ± 2.44% | 30.73 ± 2.80% | **29.37 ± 2.60%** | 30.11 ± 3.67% |
| 3 | Train_Loss | 0.6443 ± 0.1234 | 0.4627 ± 0.0907 | 0.4631 ± 0.0877 | 0.4481 ± 0.1007 | **0.3134 ± 0.1226** | 0.4358 ± 0.0980 |
|  | Validate_Loss | 0.7517 ± 0.1300 | 0.5864 ± 0.0587 | 0.5868 ± 0.0600 | 0.5798 ± 0.0638 | 0.6156 ± 0.0483 | **0.5390 ± 0.0708** |
|  | Test_Loss | 0.6389 ± 0.1119 | 0.5280 ± 0.0476 | 0.5224 ± 0.0464 | 0.5167 ± 0.0581 | 0.5757 ± 0.0340 | **0.4933 ± 0.0584** |
|  | RMSE | 50.70 ± 4.57 | 46.19 ± 2.04 | 46.06 ± 2.03 | 45.71 ± 2.38 | 47.97 ± 1.67 | **44.36 ± 2.85** |
|  | MAE | 41.89 ± 4.22 | 37.97 ± 1.94 | 37.94 ± 1.92 | 37.46 ± 2.29 | 39.03 ± 1.65 | **36.32 ± 2.81** |
|  | CSI | 52.44 ± 3.81% | 58.85 ± 2.36% | 59.16 ± 1.87% | 58.98 ± 2.47% | 58.84 ± 1.60% | **60.57 ± 2.78%** |
|  | POD | 74.16 ± 7.25% | 81.03 ± 3.14% | 83.32 ± 1.95% | 81.92 ± 2.91% | 79.40 ± 1.71% | **83.94 ± 3.34%** |
|  | FAR | 35.25 ± 5.32% | 31.71 ± 2.38% | 32.86 ± 2.37% | 32.18 ± 2.36% | **30.51 ± 2.28%** | 31.37 ± 3.63% |

For each model, their performances are calculated by averaging each 24 metrics from the total 184 cities, i.e., predicting the concentration per city per 3 hours within 72 hours in total. We repeat 10 times of experiments, and report the mean and standard deviation of metrics in Table 4 [9].

## 4.4 Experimental Results

We first present the results from automatic metrics, and then demonstrate the characteristics of the proposed model PM$_{2.5}$-GNN through careful analysis.

**Overall Performance**. As shown in Table 4, MLP, LSTM and GRU (the first three columns) perform the worst. It is not surprising because they are the models that have access to only node representations. Their disappointing performances empirically support the necessity of neighboring information for PM$_{2.5}$ prediction.

Since all other three models are enabling of capturing both temporal and spatial dependencies, we then compare them to find which method utilizes domain knowledge more effectively. Although GC-LSTM and nodesFC-GRU achieve similar scores on these automatic metrics, nodesFC-GRU has the lowest train loss and relatively higher test loss. It is a typical overfitting phenomenon when comparing nodesFC-GRU with PM$_{2.5}$-GNN and GC-LSTM in terms of parameter size.

Among all, our PM$_{2.5}$-GNN surpasses the compared models on almost every metric across the sub-datasets. Notably, it enjoys the best test loss, RMSE, MAE, CSI and POD on all the 3 datasets. This result demonstrates PM$_{2.5}$-GNN's advantage on exploiting domain knowledge and suggests its reliable prediction ability.

**Analysis on Fine-grained Scale**. As complementary for automatic evaluations, we also conduct case studies in fine-grained spatial- and temporal- scale. On Dataset 3, we compare the model predictions with the ground-truth concentrations in certain regions. This further analysis is aimed to understand why the proposed approach is effective and to what extent it is better. Without loss of generality, we select two cases. The first case is a group of four distinct regions, whose PM$_{2.5}$ distributions are depicted in Figure 7. The second case is a representative city, Xi'an.

For the first case, we adopt RMSE and Pearson correlation coefficient to quantitatively measure the differences between the predictions and the ground-truths. Given these regions, the performances in Figure 6 correlate well with those findings observed from Table 4.

---
[9] Please find the latest results in our ACM SIGSPATIAL 2020 poster paper and Github code.

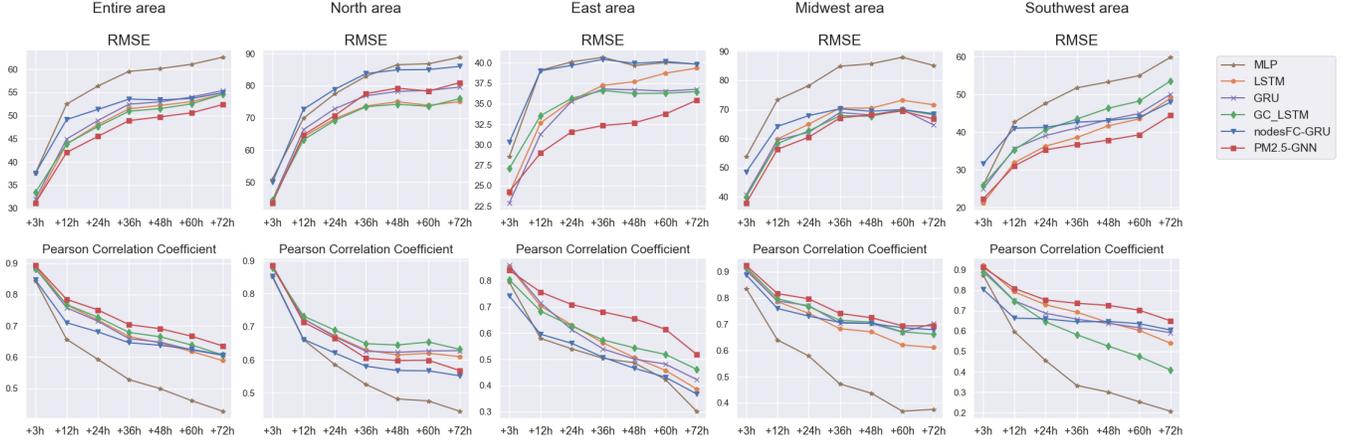

Figure 6: Performance statistics of PM$_{2.5}$-GNN and its baselines on the studied area and 4 sub-regions on testset of dataset 3 with $\tau$ hours ahead prediction, where $\tau \in [3, 12, 24, 36, 48, 60, 72]$

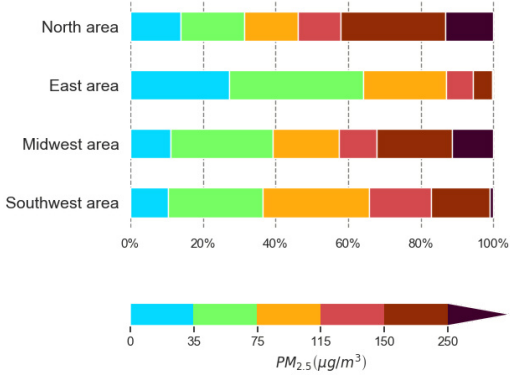

Figure 7: PM$_{2.5}$ distribution of 4 regions test set of data set 3.

Although MLP (the brown line) performs well in the first 3 hours (the first prediction step), it lags far soon. The lack of memory prohibits its modeling the PM$_{2.5}$. On the other hand, the memory cells in LSTM and GRU facilitate them to capture temporal dependencies, leading to better performances. Things become interesting in the East China (the middle figures), where GC-LSTM (the green line) and PM$_{2.5}$-GNN (the red line) present the best performances, while nodesFC-GRU (the blue line) is the worst. As indicated in Figure 7, East China has the lowest PM$_{2.5}$, whose primary cause is the transport from North China [23]. Given the observations on the blue curves, it seems that nodesFC-GRU is unsuitable for this case. It is because that the fully connection module in nodesFC-GRU introduces too many dependencies, while the transport pathways in East China are relatively few and concrete. For the case of Xi'an, we obtain similar observations. Due to limit space, please refer to Appendix to see the prediction curves.

Drawing on the best fits at every predicting phase achieved by PM$_{2.5}$-GNN, we safely conclude that our model is powerful even in extreme cases, since it: (1) takes full advantage of domain knowledge; and (2) possesses ability of modeling long-term dependencies.

Table 5: Experimental results of PM$_{2.5}$-GNN's different configurations. Lack of PBL feature or subtraction component worsens PM$_{2.5}$-GNN's performance.

| Dataset | Metric | PM$_{2.5}$-GNN | no PBL height | no export |
|---|---|---|---|---|
| 1 | RMSE | **20.16 ± 0.48** | 20.46 ± 0.43 | 20.98 ± 0.33 |
|   | MAE  | **15.91 ± 0.49** | 16.12 ± 0.44 | 16.67 ± 0.35 |
|   | CSI  | **47.91 ± 1.65%** | 46.70 ± 1.48% | 45.41 ± 1.17% |
| 2 | RMSE | **32.11 ± 1.47** | 33.25 ± 1.65 | 32.70 ± 1.31 |
|   | MAE  | **25.68 ± 1.42** | 26.67 ± 1.59 | 26.16 ± 1.27 |
|   | CSI  | **51.35 ± 2.53%** | 49.42 ± 2.90% | 50.41 ± 2.43% |
| 3 | RMSE | **44.36 ± 2.85** | 46.12 ± 3.38 | 44.80 ± 2.59 |
|   | MAE  | **36.32 ± 2.81** | 38.04 ± 3.38 | 36.78 ± 2.53 |
|   | CSI  | **60.57 ± 2.78%** | 58.72 ± 3.15% | 60.12 ± 2.38% |

**Importance of Domain Knowledge**. At last, we analyze PM$_{2.5}$-GNN from the perspective of domain knowledge. We focus on PBL height and transport direction information. These two kinds of domain knowledge are primary for PM$_{2.5}$ prediction [12, 18] but have been less explored before. Recall that the proposed PM$_{2.5}$-GNN encodes PBL height as node features (in Table 1) and captures directed transport through the knowledge-enhanced GNN component (Equation 7). Excluding the corresponding node feature and the export influence (remove $e^t_{i \to j}$ from Equation 7), we obtain the ablated results and present them in Table 5. Clearly, the performances degrade a lot when removing any of them. This ablation study verifies the importance of domain knowledge as well as the motivation of our research.

## 5 CONCLUSIONS

In this paper, we study a significant problem in real-world that how to precisely predict PM$_{2.5}$ concentrations in the next 72 hours. Observing the two typical PM$_{2.5}$ characteristics, we leverage a GNN to introduce domain knowledge and integrate with a RNN to capture the fine-grained and long-term dependencies during the PM$_{2.5}$ process. We show the success of our approach through extensive experiments and deploy the proposed PM$_{2.5}$-GNN model online in hope to benefit the community. In the future, we plan to study the model interpretability and PM$_{2.5}$ source tracking. It is also promising to generalize our approach to PM$_{10}$ (dust) concentration prediction which has more obvious transport effect.

## ACKNOWLEDGMENTS

This research is supported by the National Natural Science Foundation of China (NSFC) under the grant number 61673070.

## A  DEPLOYMENT DETAILS

The model $PM_{2.5}$-GNN described in this paper has been deployed online through several service channels, including air forecasting website, air forecasting mobile Apps and API protocols for third-party clients. The deployment architecture is illustrated in Figure 10.

Compared to offline settings, there are two notable differences when we deploy our model online. First of all, we use different weather data sources to fit different settings. During offline experiments, we we obtain climate reanalysis ERA5[10] from European Centre for Medium-Range Weather Forecasts (ECMWF)[11] as the weather forecasting data. This weather data is re-analyzed and fixed using real historical data, which is not real-time available, and in turn is unable to be used for online service. To this end, for online deployment, we acquire weather data from Global Forecast System (GFS)[12] since it is charge-free. GFS publishes its forecasting 4 times a day with few time delays, which contains future 15-days weather data with detailed information (as introduced in Section 3.3).

The other difference resides in the deployment framework. As illustrated in Figure 10, we read raw data from Ministry of Ecology and Environment of China (MEE) and Global Forecast System (GFS), and store them into a spatio-temporal database for online use. Because $PM_{2.5}$ concentration is sensitive to meteorological information and other relative domain knowledge, we update the raw data everyday and re-train our $PM_{2.5}$-GNN model every midnight in online settings. Assuming that $PM_{2.5}$ concentration distributions often remain stable within three months, we train and update our online $PM_{2.5}$-GNN model everyday using data of the past three months. Then, the updated model is deployed on the server to provide $PM_{2.5}$ predictions to clients through APIs.

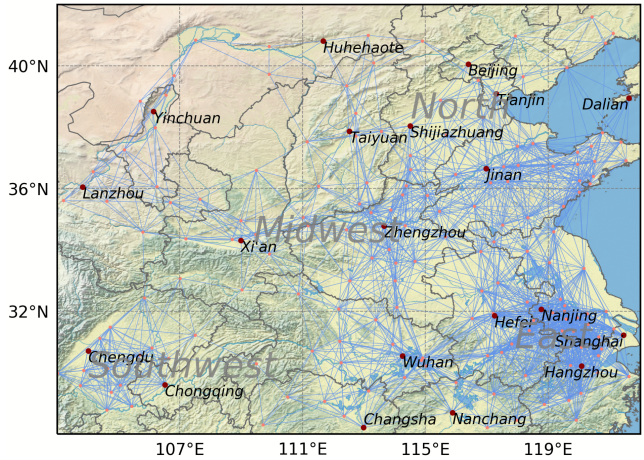

Figure 8: The Studied Area Shown On the Map.

## B  STUDIED AREA

In this work, we introduce a large-scale real-world dataset **KnowAir**, which includes a wide range of areas for $PM_{2.5}$ forecasting. These areas range from $103°E - 122°E$ and $28°N - 42°N$, and covers several severely polluted regions in China.

[10]https://climate.copernicus.eu/climate-reanalysis
[11]https://www.ecmwf.int/
[12]https://www.ncdc.noaa.gov/data-access/model-data/model-datasets/global-forcast-system-gfs

In Figure 8, we show the studied areas on map. The dots denote the cities and blue lines represent their potential interactions: $PM_{2.5}$ has possibility to be transported from one to another city when the distance between these cities is shorter than a threshold and there does not exist high mountains to obstacle the transport..

Note the studied areas in our dataset KnowAir is far large from those in previous work [16].

## C  $PM_{2.5}$-GNN PROCEDURE

We summarize the learning procedure of the proposed $PM_{2.5}$-GNN in below.

---

**Algorithm 1:** $PM_{2.5}$-GNN model

**Input** : $PM_{2.5}$ observed concentrations $X^0$ ;
  nodes' attributes $[P^1, ..., P^T]$;
  edges' attributes $[Q^1, ..., Q^T]$;
  $G = (V, E)$;
**Output:** $PM_{2.5}$ predicted concentrations $[\hat{X}^1, ..., \hat{X}^T]$;

#Initialize:
$h^0 \leftarrow 0$;
$\hat{X}^0 \leftarrow X^0$;
$Output\_list = [\ \ ]$;

**for** $t = 1, ..., T$ **do**
  **for** $i \in V$ **do**
    $\zeta_i^t = \mathbf{GNN}(\xi_i^t, \{\xi_j^t, Q_{i \to j}^t, Q_{j \to i}^t\}_{j \in N(i)})$ (Equation 7);
    $h_i^t = \mathbf{GRUcell}([\xi_i^t, \zeta_i^t], h_i^{t-1})$ (Equation 8);
    $\hat{X}_i^t = \mathbf{MLP}(h_i^t)$ (Equation 9);
    Append $\hat{X}_i^t$ into $Output\_list$;

---

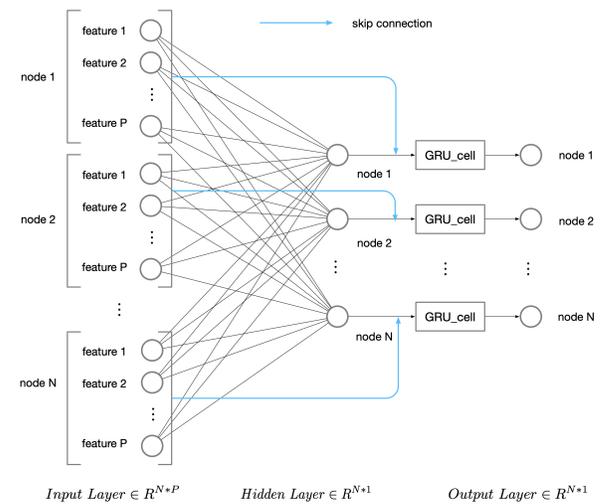

Figure 9: nodesFC-GRU

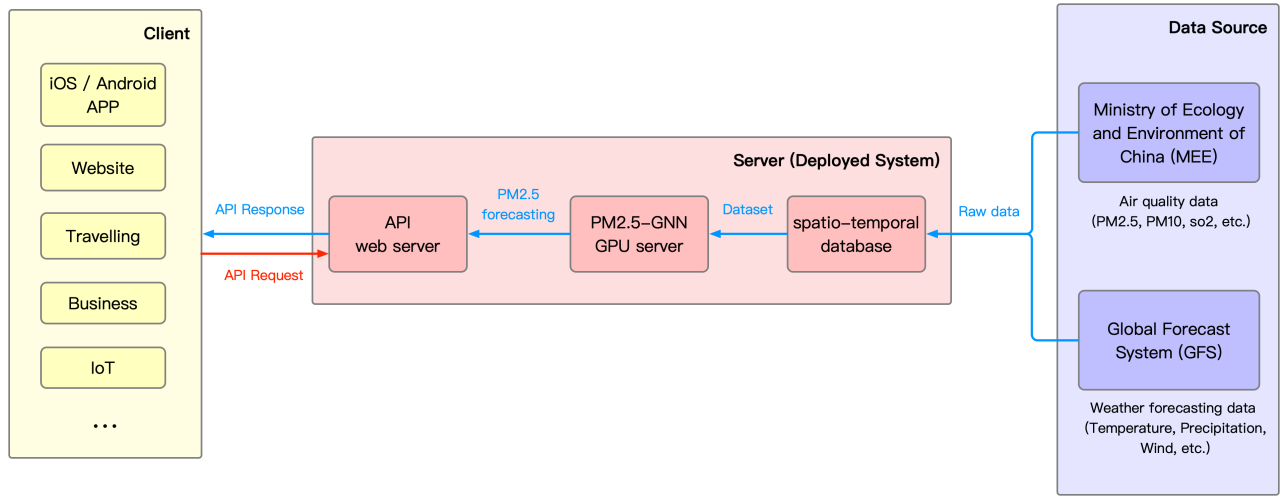

Figure 10: Deployment Framework

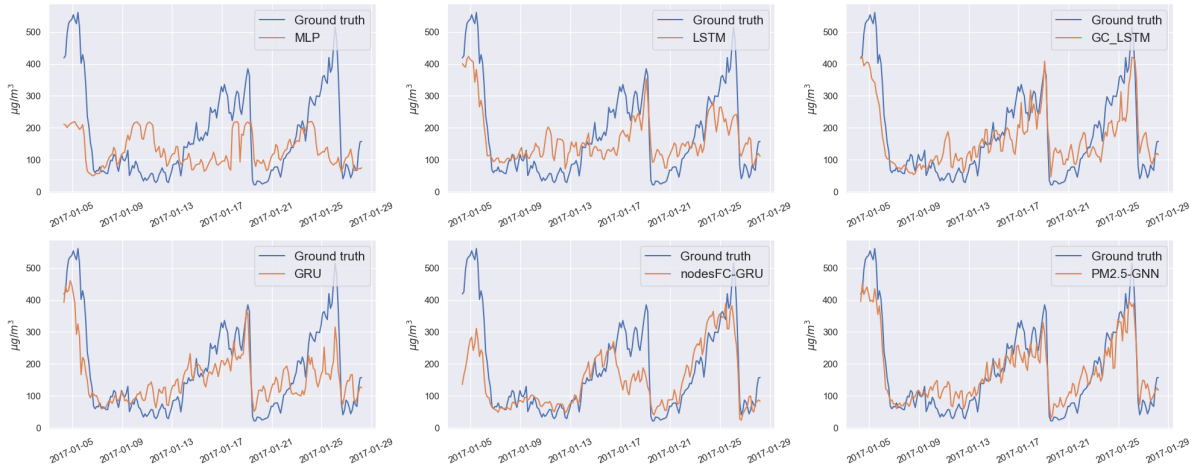

Figure 11: Comparison of prediction curves between PM$_{2.5}$-GNN and its baselines for 72 hours ahead prediction at Xi'an node of testset of Dataset 3.

## D  NODESFC-GRU

The proposed is PM$_{2.5}$-GNN consists of two main components: A knowledge-enhanced Graph Neural Network (GNN) and a spatio-temporal GRU.

To study the significance of GNN in the proposed approach, we replace the GNN module in PM$_{2.5}$-GNN with a fully connected (FC) MLP. This degrading version is named as **nodesFC-GRU**. By comparing with it, we aim to examine how much improvement can be brought by neighborhood information using such a naive method.

In Figure 9, we illustrate the architecture details of nodesFC-GRU. According to the experimental results from Table 4 and Figure 6, it is undesirable to predict PM$_{2.5}$ concentration using nodesFC-GRU especially when PM$_{2.5}$ factors only put on a relatively weak effect on the considered place. Under such situations, the large amount of parameters in the fully connection module inevitably leads nodesFC-GRU to overfit the data.

## E  CASE STUDY: XI'AN

As complementary for automatic evaluations, we select a typical city, Xi'an (from Dataset 3) to conduct a case study in fine-grained scale.

In Figure 11, we depict the predictions from six models and the ground-truth concentrations. It is clear that PM$_{2.5}$-GNN's prediction curve fits best to the ground-truth curve.

As a summary of the results from both automatic evaluation and case studies, the proposed PM$_{2.5}$-GNN predicts PM$_{2.5}$ more accurately by leveraging extensive domain knowledge effectively.